# Topology hierarchy of transition metal dichalcogenides built from quantum spin Hall layers

Lixuan Xu, Yiwei Li*, Yuqiang Fang, Huijun Zheng, Wujun Shi, Cheng Chen, Ding Pei, Donghui Lu, Makoto Hashimoto, Meixiao Wang, Lexian Yang, Xiao Feng, Haijun Zhang, Fuqiang Huang, Qikun Xue, Ke He*, Zhongkai Liu*, and Yulin Chen*

L. Xu, L. Yang, X. Feng, Q. Xue, K. He
State Key Laboratory of Low Dimensional Quantum Physics
Department of Physics
Tsinghua University
Beijing 100084, China
Email: kehe@tsinghua.edu.cn

Y. Li
Institute for Advanced Studies (IAS)
Wuhan University
Wuhan 430072, China
Email: yiweili@whu.edu.cn

H. Zheng, W. Shi, D. Pei, M. Wang, Z. Liu, Y. Chen
School of Physical Science and Technology
ShanghaiTech University
Shanghai 201210, China
Email: liuzhk@shanghaitech.edu.cn; yulin.chen@physics.ox.ac.uk

Y. Fang, F. Huang
State Key Laboratory of Rare Earth Materials Chemistry and Applications College of Chemistry and Molecular Engineering
Peking University
Beijing 100871, China

H. Zheng, D. Pei, M. Wang, Z. Liu, Y. Chen
ShanghaiTech Laboratory for Topological Physics
Shanghai 201210, China

W. Shi
Center for Transformative Science
ShanghaiTech University
Shanghai 201210, China

W. Shi
Shanghai high repetition rate XFEL and extreme light facility (SHINE)
ShanghaiTech University
Shanghai 201210, China

C. Chen
Advanced Light Source
Lawrence Berkeley National Laboratory
Berkeley, CA 94720, USA

D. Lu, M. Hashimoto
Stanford Synchrotron Radiation Lightsource
SLAC National Accelerator Laboratory
Menlo Park, California 94025, USA

H. Zhang
National Laboratory of Solid State Microstructures and School of Physics
Nanjing University
Nanjing 210093, China

Y. Chen
Clarendon LaboratoryDepartment of Physics
University of Oxford
Oxford OX1 3PU, UK

# Abstract

The evolution of the physical properties of two-dimensional material from monolayer limit to the bulk reveals unique consequences from dimension confinement and provides a distinct tuning knob for applications. Monolayer 1T'-phase transition metal dichalcogenides (1T'-TMDs) with ubiquitous quantum spin Hall (QSH) states are ideal two-dimensional building blocks of various three-dimensional topological phases. However, the stacking geometry was previously limited to the bulk 1T'-WTe$_2$ type. Here, we introduce the novel 2M-TMDs consisting of translationally stacked 1T'-monolayers as promising material platforms with tunable inverted bandgaps and interlayer coupling. By performing advanced polarization-dependent angle-resolved photoemission spectroscopy as well as first-principles calculations on the electronic structure of 2M-TMDs, we revealed a topology hierarchy: 2M-WSe$_2$, MoS$_2$, and MoSe$_2$ are weak topological insulators (WTIs), whereas 2M-WS$_2$ is a strong topological insulator (STI). Further demonstration of topological phase transitions by tunning interlayer distance indicates that band inversion amplitude and interlayer coupling jointly determine different topological states in 2M-TMDs. We propose that 2M-TMDs are parent compounds of various exotic phases including topological superconductors and promise great application potentials in quantum electronics due to their flexibility in patterning with two-dimensional materials.

# 1. Introduction

The last decade has witnessed a tremendous triumph in two-dimensional (2D) materials which host intriguing properties such as high mobility electrons, unusual superconductivity, excitonic behaviors, 2D ferromagnetism, as well as topological quantum physics.[1-5] The stacking of atomically thin materials acts as an additional turning knob for physical properties and application purposes as the thickness evolves from the monolayer limit to few-layer and bulk crystals.[6-18] For example, semiconductor 2H-MoSe$_2$ experiences a direct to indirect bandgap transition as the thickness increases from monolayer to bilayer.[6] MnBi$_2$Te$_4$ can realize quantum anomalous Hall insulators, axion insulators, and antiferromagnetic topological insulators depending on the thickness and even/odd layer numbers.[7-13] Recent studies revealed rich phase diagrams of twisted multi-layer graphene systems exhibiting various exotic phases including correlated insulating phases, superconductivity, and topological phases as functions of twist angles, filling numbers, and displacement fields.[14-18]

The topological phase transition by stacking 2D materials has become a central topic of topological quantum materials, e.g. the topological hierarchy by stacking the quantum spin Hall (QSH) layers. QSH states are unique 2D systems featuring topologically protected helical edge states.[19] Despite their promising application potential in low-dissipation quantum devices, the theoretical proposals and experimental realization of QSH states are limited to very few materials, such as semiconductor quantum wells and buckled two-dimensional Xene sheets.[20-22] Among them, monolayer 1T'-phase transition metal dichalcogenides (1T'-TMDs) stand out as this material family host ubiquitous QSH states[23] and they are evidenced by both spectroscopic and transport experiments.[22,24,25] Three-dimensional (3D) crystals built from

1T'-TMD monolayers are previously limited to WTe$_2$ and MoTe$_2$ hosting alternative stacking of flipped 1T'-monolayers, achieving higher-order topological insulator (HOTI) phase[26] (1T'-WTe$_2$ and 1T'-MoTe$_2$, see Figure 1a-ii) and Weyl semimetal (WSM) phase[27,28] (slightly distorted Td-WTe$_2$ and Td-MoTe$_2$).

In this work, we introduce the novel 2M-TMDs consisting of translationally stacked 1T'-monolayers[29,30] and demonstrate that they are ideal material platforms for studying topological phases of stacked QSH layers[31,32] with tunable inverted bandgaps and interlayer coupling strengths (see Figure 1a). Combining first-principles calculations and angle-resolved photoemission spectroscopy (ARPES), we systematically investigated the electronic structure of 2M-TMDs, revealing a topology hierarchy. We discovered that 2M-WSe$_2$, MoSe$_2,$ and MoS$_2$ host weak topological insulator (WTI) phases which can be regarded as weakly coupled QSH layers (see Figure 1a-iii), whereas 2M-WS$_2$ is a strong topological insulator (STI) exhibiting topological surface states (TSSs) on the naturally cleaved surface (see Figure 1a-iv). By calculating the electronic structure evolution with variable interlayer distance, we demonstrate that a topological phase transition from the STI phase to the WTI phase can be achieved by a weakened interlayer coupling. Similar concepts have recently been reported in quasi-one-dimensional bismuth halide system[33-35] and TaSe$_3$.[36,37] More importantly, recent intriguing discoveries in 2M-TMDs, such as the coexistence of TSSs and superconductivity,[29,38,39] signatures of zero Majorana modes[40] and Pauli-limit violated superconductivity,[41] make them not only ideal for fundamental research, but also promising materials for novel applications.

## 2. Experimental Results and Discussion

2M-TMDs crystallize in a base-centered monoclinic structure (space group $C_{2/m}$, No. 12), as shown in Figure 1b-i with the conventional and primitive unit cell indicated by the black and magenta parallelepiped, respectively. Within each layer (defined as *b-c* or *x-y* plane), it has the 1T'-structure (see Figures 1b-ii to iv). The van der Waals layers of 2M-TMDs, including 2M-WS$_2$, WSe$_2$, MoSe$_2$, and MoS$_2$, are stacked via a direct translation vector[29,30,38,40] (defined as the *a* axis of the primitive cell marked by the magenta arrow in Figure 1b), in contrast to the glide mirror stacking geometry (a combination of translation and mirror operations) in the bulk 1T'-WTe$_2$[42] and 1T'-MoTe$_2$[28] (a comparison of the crystal structures between the bulk 2M- and 1T'-TMDs can be found in Supporting Information Note 1).

We performed first-principles calculations on the electronic structures of all four group VIB 2M-TMDs. Figure 2a presents typical band structures of 2M-WSe$_2$ and 2M-WS$_2$ as examples (the results of the other two compounds can be found in Supporting Information Note 6). All four materials are semimetallic in contrast to their semiconducting 2H polymorphs, however, the top valence band and the bottom conduction band are well defined with a non-zero bandgap at any arbitrary momenta (the gray-shaded areas in Figure 2a).

Orbital analysis indicates that the inverted band structures near Γ in the monolayer 1T'-TMDs[23] preserve in all four 2M-TMDs, showing that the top valence band mainly consists of the transition metal *d*-orbitals and the bottom conduction band mainly consists of chalcogenide *p*-orbitals (see Figure 2a). Notably, the band orbital ordering changes along the ΓZ direction in 2M-WS$_2$ (indicated by the red arrow in the lower panel of Figure 2a) due to the interlayer interaction, whereas in 2M-WSe$_2$, the inverted band ordering preserves from Γ to Z (see the upper panel of Figure 2a). Hence, they present two classes of topological band structures of stacked QSH layers as described in Figure 2b: 2M-WSe$_2$ is a WTI where the band inversions occur at both Γ and Z, and in comparison, 2M-WS$_2$ is a STI where the band inversion only occurs at Γ since the interlayer coupling is strong enough to "untie" the inverted band order at Z. The bandgap at Γ (Δ(Γ)) and Z (Δ(Z)) of all four 2M-TMDs, as well as the inverted bandgap of the monolayer 1T'-TMD (Δ(mono)), are summarized in Figure 2c). One can see that 2M-WS$_2$ hosts the smallest Δ(Γ), Δ(mono) and it is the only compound with a positive Δ(Z) among the four materials, leading to a STI phase whereas the other three materials are WTIs.

The topology hierarchy is further checked by calculating the topological invariant $Z_2$ = ($v_0$; $v_1v_2v_3$). Since 2M-TMDs are centrosymmetric, we calculated the product of the parities of the Bloch wavefunction for the occupied bands at all eight time-reversal-invariant-momenta in the bulk Brillouin zone (as indicated by the red spheres in Figure 2d) based on the method proposed by Fu and Kane.[43] Figure 2e shows the calculated parity products and the corresponding $Z_2$ numbers of these four materials, consistent with the orbital analysis.

We carried out a systematic electronic structure investigation on the WTI candidate 2M-WSe$_2$ and STI candidate 2M-WS$_2$ by ARPES (detailed methods can be found in Experimental Section), as presented in Figure 3. 2M-WSe$_2$ hosts a pair of hole pockets along the $\bar{Y} - \bar{\Gamma} - \bar{Y}$ direction (see Figures 3a,b). The ARPES measured band dispersions along two perpendicular high-symmetry momentum directions $\bar{Y} - \bar{\Gamma} - \bar{Y}$ and $\bar{X} - \bar{\Gamma} - \bar{X}$ show excellent agreement with our first-principles calculations (see Figure 3c). The valence band near $\bar{\Gamma}$ along the $\bar{Y} - \bar{\Gamma} - \bar{Y}$ direction exhibit a "W" shape (see the left panel of Figure 3c), indicative of the band inversion at $\bar{\Gamma}$. We observed no sign of the surface Dirac cone at $\bar{\Gamma}$ crossing the Fermi level (see Figures 3b,c and Supporting Information Note 9), consistent with the WTI case as the (100) cleavage surface hosts no TSSs.

In comparison, ARPES measurements on the STI candidate 2M-WS$_2$ show its Fermi surfaces with rich details that can be successfully reproduced by first-principles calculations (see Figures 3d,e), consistent with our previous study.[38] Using laser-based ARPES with high momentum and energy resolution ($\Delta k$ = 0.003 Å$^{-1}$, $\Delta$E = 3 meV), we observed the signature of the expected surface Dirac point in the fine Fermi surface mapping, as shown in the zoom-in momentum area near $\bar{\Gamma}$ (see Figure 3e-ii). The TSSs form the surface Dirac cone at $\bar{\Gamma}$ and were clearly seen in the band dispersion measurement along the $\bar{Y} - \bar{\Gamma} - \bar{Y}$ direction (Figures 3f-i,ii), again consistent with the calculations (Figures 3f-iii,iv).

The tunable excitation energy ($h\nu$) and polarization of incident photons in the synchrotron-based ARPES measurements (see Figure 4a) enable us to discriminate band orbitals with different symmetries along the ΓZ direction.[44] Using photon energies of 84 eV and 99 eV, we measured the band dispersion along $\bar{Y} - \Gamma - \bar{Y}$ (see Figure 4b-i) and $\bar{Y} - Z - \bar{Y}$ (see Figure 4b-ii), respectively, which is consistent with corresponding calculations. According to the matrix element effect, orbitals of odd (even) symmetry with respect to the mirror plane defined by the incident light and emission electron are visible under linear vertical (linear horizontal) polarized light with the electric field out of (in) the mirror plane (see detailed discussion in Supporting Information Note 3). For both measurements near Γ and Z, the first and second topmost valence bands show significant spectral weight under linear horizontal (LH) and linear vertical (LV) polarized light, respectively (see Figures 4b-iii to vi). These observations agree with the orbital-resolved calculation indicating that the two topmost valence bands near Γ are mainly from $W$-$d_{xz}$ and $Se$-$p_z$ orbitals whereas near Z, they mainly consist of $W$-$d_{xz}$ and $W$-$d_{z2}$

(detailed first-principles calculations on the orbital-resolved band structure can be found in Supporting Information Note 4). The photon-energy and polarization selective ARPES results thus provide direct experimental evidence on the orbital nature of the band structure of 2M-WSe$_2$, indicative of the WTI scenario exhibiting band inversions at both Γ and Z as shown in Figure 2b.

Band dispersions along the $k_z$ direction of 2M-WS$_2$ and 2M-WSe$_2$ are measured via photon-energy-dependent ARPES, as shown in Figure 4c (details of the photon-energy-dependent measurement can be found in Supporting Information Note 2). In 2M-WS$_2$, due to the interlayer coupling, the band dispersion along Γ-Z is strong enough to induce the band order switching between the topmost valence and the lowest conduction band. In comparison, the topmost valence band of 2M-WSe$_2$ shows negligible $k_z$ dispersion. The distinct $k_z$ band evolutions of 2M-WS$_2$ and 2M-WSe$_2$ are consistent with the STI and WTI scenario illustrated in Figure 2b.

Combining first-principles calculations and systematic ARPES investigation, we provide 2M-WSe$_2$ and 2M-WS$_2$ as two typical examples of stacked QSH layers with distinct topological phases. To fully understand the electronic structure of 2M-TMDs and their topological origins, first-principles calculations were performed on 2M-WS$_2$ demonstrating that the band structure of the STI phase can evolve to the WTI phase by continuously increasing the interlayer distance (see Figure 5). Since 2M-WSe$_2$ is at the WTI end, a WTI to STI topological phase transition can be triggered by continuously decreasing the interlayer distance (see Supporting Information Note 8).

The bulk electronic structure evolution (see Figure 5a) nicely illustrates the mechanism of the topological phase transition from the STI phase to the WTI phase induced by the additional band inversion at Z with increasing interlayer distance and thus weakening dispersion along the Γ-Z direction, which are manifested as the closing and reopening of a bandgap between the valence and conduction bands with opposite parities. The orbital character of *W-d* and *S-p* orbitals at Z switches between the conduction and valence bands through the phase transition, as clearly presented in Figure 5e. The surface projected band dispersions (see Figures 5b,c) exhibit expected evolutions hosting the TSSs inside the inverted bandgap in the STI region, whereas no TSS exists on the (100) surface in the WTI region. One can see that the surface Dirac point gradually merges into the bulk states and becomes the bulk Dirac point, suggestive of an unstable Dirac semimetal (DSM) phase at the critical point ($\Delta a \approx 2$Å) of the topological phase transition between the STI phase and the WTI phase (see Figure 5f) as has been theoretically modeled.[45]

We calculated the energy of 2M-WS$_2$ as a function of the interlayer distance, as shown in Figure 5d. Due to its van der Waals layered structure, the critical point of the phase transition is about 0.1 eV/f. u above the equilibrium energy. In this case, multiple experimental approaches might trigger the topological phase transition, such as elemental substitution, intercalation, and laser excitation, which promises a rich material playground for the study of topological phases with tunable parameters.[46-48]

## 3. Summary and Outlook

Bulk topological materials consisting of QSH layers have been theoretically proposed and experimentally realized in various compounds. Experimental signatures of topological edge states have been observed on the bulk crystal stepping edges of WTe$_2$,[49] transition-metal pentatelluride ZrTe$_5$ and HfTe$_5$,[50,51] bismuth halide Bi$_4$I$_4$ and Bi$_4$Br$_4$,[33-35, 52,53] transition-metal tritelluride TaSe$_3$,[36,37] and CaSn,[54] however, experimental confirmations of the QSH states in the monolayer forms are limited to very few materials.[55] 2M-TMD family and their sister compounds WTe$_2$ and MoTe$_2$ stand out for the following reasons. Firstly, although monolayer 1T'-TMDs share similar electronic structures hosting ubiquitous and robust QSH states,[22-25] the bulk counterparts exhibit distinct topological phases in their pristine forms: 2M-WSe$_2$, MoSe$_2$, and MoS$_2$ are WTIs; 2M-WS$_2$ is a STI; 1T'-WTe$_2$ and 1T'-MoTe$_2$ are HOTIs; Td-MoTe$_2$ is a WSM (see Figure 1a). Secondly, the flexibility of patterning with other semiconducting TMDs could have technological applications in novel quantum electronics.[56] We propose that pressure-dependent[57] or strain-controlled measurements[36,51] on the 2D-heoterostructure including 2M-TMDs might provide potential dynamic control of QSH edge states by effective tunning of interlayer coupling strength. Thirdly, rich superconducting phase diagrams of 2M-TMDs as functions of chemical compositions, pressure, and carrier density imply this material family as a promising material platform for the study of topological superconductors.[58-61] Hence, we provide a systematic investigation of the electronic structure of 2M-TMD family combining first-principles calculations and ARPES measurements, which will motivate further fundamental and technological research for topological materials and the new generation of low-dissipation quantum devices, especially topological superconductors and quantum computing based on QSH layers.

## 4. Experimental Section

***Sample synthesis:*** 2M-WS$_2$ single crystals were prepared by the deintercalation of interlayer potassium cations from K$_{0.7}$WS$_2$ crystals. For the synthesis of K$_{0.7}$WS$_2$, K$_2$S$_2$ (prepared via liquid ammonia), W (99.9%, Alfa Aesar) and S (99.9%, Alfa Aesar) were mixed by the stoichiometric ratios and ground in an argon-filled glovebox. The mixtures were pressed into a pellet and sealed in the evacuated quartz tube. The tube was e was heated at 850 °C for 2000 min and slowly cooled to 550 °C at a rate of 0.1 °C/min. The synthesized K$_{0.7}$WS$_2$ (0.1 g) was oxidized chemically by K$_2$Cr$_2$O$_7$ (0.01 mol/L) in aqueous H$_2$SO$_4$ (50 mL, 0.02 mol/L) at room temperature for 1 h. Finally, the 2M-WS$_2$ crystals were obtained after washing in distilled water for several times and drying in the vacuum oven at room temperature.[29]

Similarly, 2M-WSe$_2$ crystals were prepared by the deintercalation of interlayer potassium cations from K$_x$WSe$_2$ crystals. The K$_x$WSe$_2$ crystals were synthesized by the reaction of the K and 2H-WSe$_2$ at 800 °C. Then the K$_x$WSe$_2$ crystals were soaked in an oxidizing solution to extract K$^+$ ions to obtain 2M-WSe$_2$ crystals.[62]

***First-Principles Calculation:*** The density functional theory (DFT) calculations were carried out via the *Vienna Ab initio Simulation Package* (VASP).[63] The projector-augmented wave (PAW) method and the plane-wave basis with an energy cutoff of 400 eV were adopted. The exchange-correlation energy was approximated by the Perdew-Burke-Ernzerhof (PBE) type generalized gradient approximation (GGA).[64] Besides, the IRVSP[65] codes and p4VASP program were used for post-processing of the VASP calculated data.

The experimental lattice constants (the cif documents) were taken from Ref[30] and the structural relaxation for optimized lattice constants and atomic positions was performed with a force criterion of 0.01 eV/Å and by using the DFT-D3 method[66] to include van der Waals corrections.

Spin-orbit coupling was included in self-consistent calculations and the Monkhorst-Pack k-point mesh of 17×9×9 was adopted.

Topological properties, including surface state calculations and Z2 characterizations of bulk materials along with edge states and Chern numbers of the monolayer, performed with WannierTools package,[67] based on the tight-binding Hamiltonians constructed from maximally localized Wannier functions (MLWFs) by the *Wannier90* package.[68-70]

To fully understand the electronic structure of 2M-TMDs and their topological origins, *ab initio* calculations were performed for these materials by continuously increasing or decreasing the interlayer distance while keeping the in-plane lattice constant unchanged.

*Angle-Resolved Photoemission Spectroscopy:* Synchrotron-based ARPES data were performed at Spectromicroscopy beamline of Elettra Synchrotron, Italy, beamline BL07U of Shanghai Synchrotron Radiation Facility (SSRF), China, and beamline BL5-2 of Stanford Synchrotron Radiation Laboratory (SSRL), Stanford Linear Accelerator Center (SLAC), USA. The sample was cleaved *in situ* and aligned the $\bar{\Gamma} - \bar{Y}$ direction parallel to the analyzer slit. The measurements were carried out under ultra-high vacuum below $5\times10^{-11}$ Torr. Data were collected by an internal movable electron energy analyzer at Spectromicroscopy of Elettra Synchrotron, and Scienta DA30L analyzers at beamline BL07U of SSRF, China and beamline BL5-2 of SSRL, SLAC, USA. The total energy resolutions were below 20 meV and the angle resolution was 0.2°.

High-resolution laser-based ARPES measurements were performed at a home-built setup (*hv* = 6.994 eV) at ShanghaiTech University. The sample was cleaved *in situ* and aligned the $\bar{\Gamma} - \bar{Y}$ direction parallel to the analyzer slit. The measurements were carried out under ultra-high vacuum below $5\times10^{-11}$ Torr. Data were collected by a DA30L analyzer. The total energy and angle resolutions were 3 meV and 0.2°, respectively.

## Supporting Information

Supporting Information is available from the Wiley Online Library or from the author.

## Acknowledgements

L.X., Y.L. contributed equally to this work. Y.L. acknowledges the support from the National Natural Science Foundation of China (12104304). Z.L. acknowledges the National Key R&D program of China (Grant No. 2017YFA0305400).

## Conflict of Interest

The authors declare no conflict of interest.

## Data Availability Statement

The data that support the findings of this study are available from the corresponding author upon reasonable request.

## Keywords

transition metal dichalcogenides, quantum spin Hall, topology phase transition, angle resolved photoemission spectroscopy

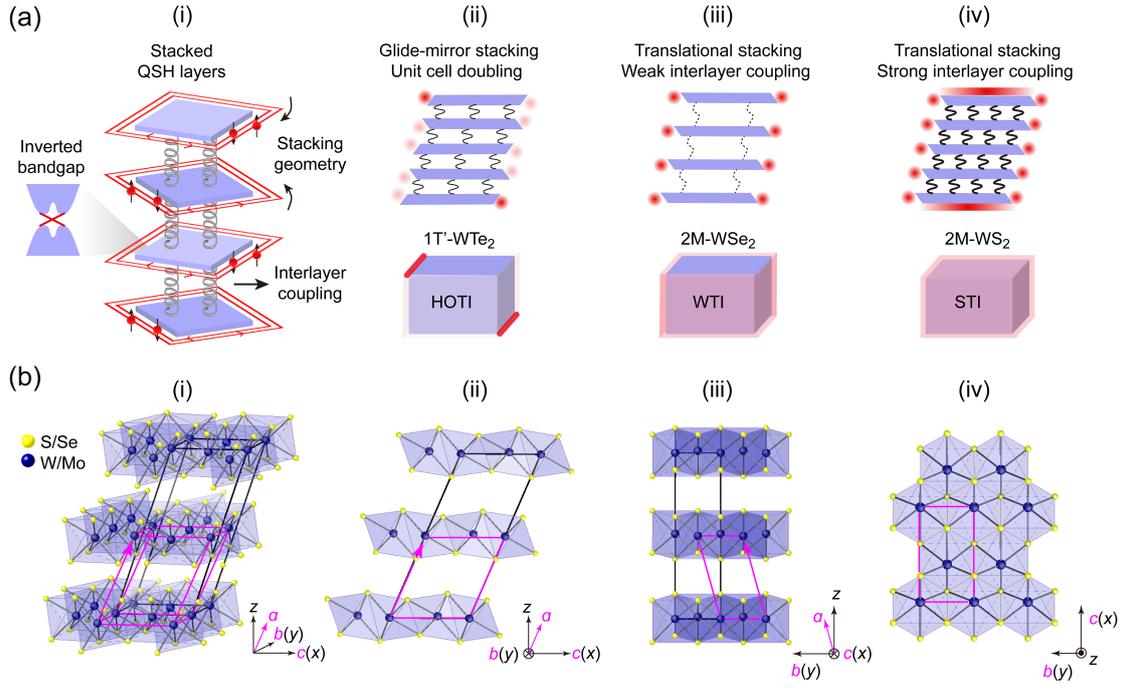

Figure 1. Transition metal dichalcogenides (TMDs) built from stacked quantum spin Hall (QSH) layers. a) Schematic plot of stacked QSH layers with tunable inverted bandgaps, stacking geometry, and interlayer coupling (i) and three material examples of stacked QSH layers exhibiting a distinct topology hierarchy (ii-iv). In (a)(ii), 1T'-WTe$_2$ is stacked by the glide mirror operation, which results in the higher-order topological insulator (HOTI) phase with gapless topological hinge states and gapped surface states. In (a)(iii), 2M-WSe$_2$ consists of weakly coupled QSH layers with translational stacking geometry, which results in the weak topological insulator (WTI) phase hosting gapless topological surface states (TSSs) on the side surfaces. In (a)(iv), 2M-WS$_2$ consists of strongly coupled QSH layers with translational stacking geometry, which results in the strong topological insulator (STI) phase hosting gapless TSSs on all surfaces. b) 3D schematic plot of the crystal structure of 2M-TMDs (i) and projected schematic plots of the crystal structure on the *xz*- (ii), *yz*- (iii) and *xy*- (iv) plane. The boundaries of the primitive and conventional unit cells are indicated by magenta and black lines, respectively.

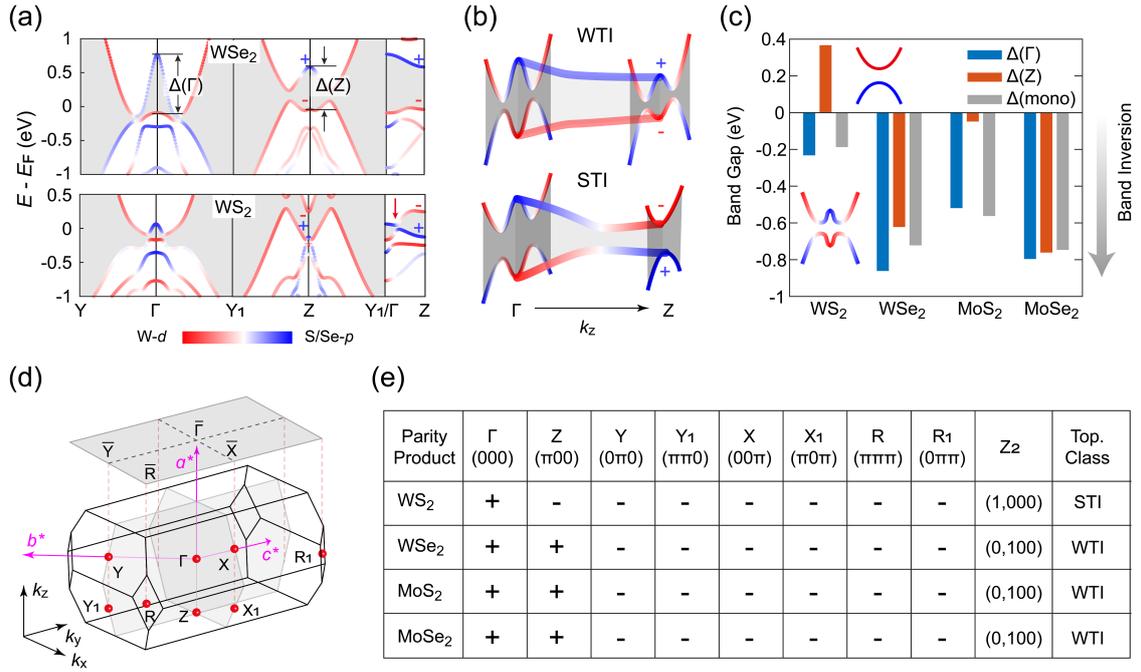

Figure 2. Band structures and topological classes of 2M-TMDs. a) Calculated orbital-resolved bulk band structure of 2M-WSe$_2$ (upper panel) and 2M-WS$_2$ (lower panel). The gray-shaded area indicates the continuous bandgap that separates the valence and conduction bands. The parity eigenvalues at Z are indicated as + (even) and – (odd). The red arrow in the lower panel denotes the switching of the orbital character along the ΓZ direction. b) 3D schematic plots of the band structure of the WTI (upper panel) and the STI (lower panel) phases constructed from QSH layers, indicating that the inverted bands are "untied" at Z in the STI case. c) Bandgap sizes at Γ (Δ(Γ)) and Z (Δ(Z)) of all four 2M-TMDs, as well as the inverted bandgap of the corresponding monolayer 1T'-TMDs (Δ(mono)). Δ(Γ) and Δ(Z) are also indicated in the upper panel of (a). Upper inset: Illustration of a normal band order with a positive band gap, which corresponds to the band structure of a STI (2M-WS$_2$) near Z. Lower inset: Illustration of an inverted band order with a negative band gap, which corresponds to the band structure of a STI (2M-WS$_2$) near Γ and a WTI (2M-WSe$_2$) near both Γ and Z. d) Bulk Brillouin zone (BZ) of 2M-TMDs with eight time-reversal-invariant-momenta marked by the red spheres. The gray rectangle shows the 2D BZ of the (100) natural cleavage surface. e) The parity products, $Z_2$ numbers, and topological classes of all four 2M-TMDs.

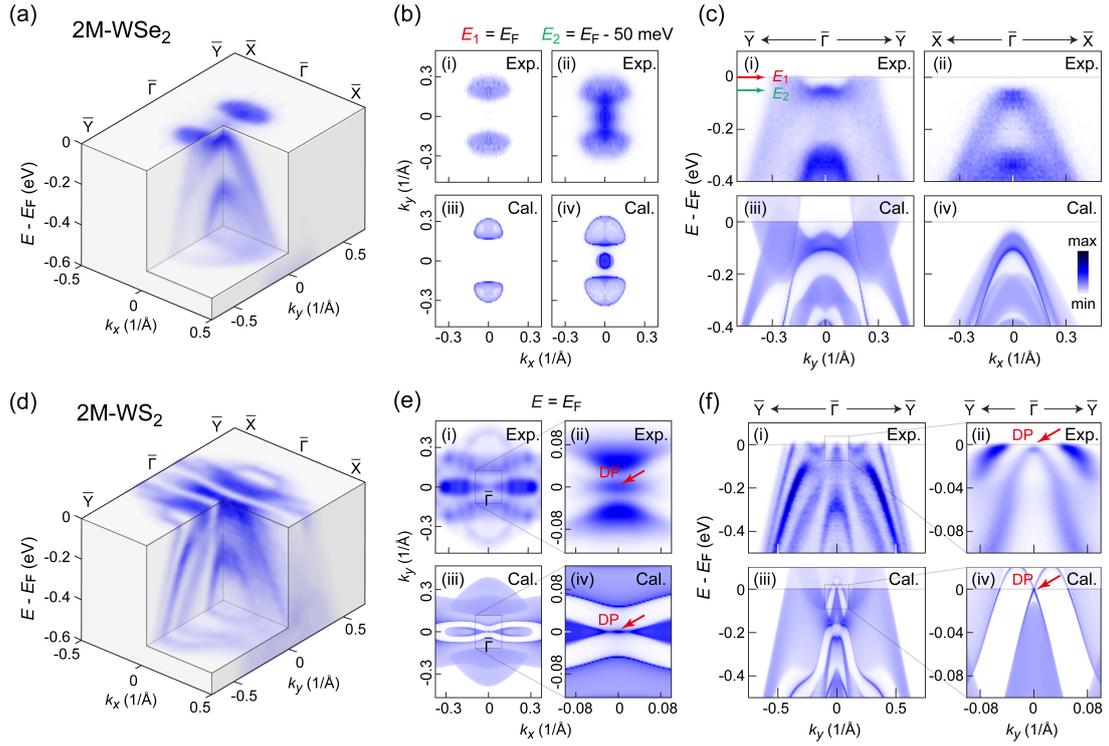

Figure 3. ARPES measurements on the electronic structures of 2M-WSe$_2$ and 2M-WS$_2$. a) A 3D plot of the photoemission spectra on the (100) surface of 2M-WSe$_2$. b) Comparisons showing nice agreement between experiments (i,ii) and first-principles calculations (iii,iv) of two constant energy contours at $E_F$ and $E_F - 50$ meV of 2M-WSe$_2$. $k_z$ integration window of the calculation (iii,iv) is ±0.2 Å$^{-1}$ around the Z-plane. c) Comparisons between photoemission dispersions (i,ii) and corresponding calculations (iii,iv) cutting along $\bar{Y} - \bar{\Gamma} - \bar{Y}$ and $\bar{X} - \bar{\Gamma} - \bar{X}$ of 2M-WSe$_2$. d) A 3D plot of the photoemission spectra on the (100) surface of 2M-WS$_2$. e) Fermi surface mapping of 2M-WS$_2$ measured by synchrotron-based ARPES (i) and a zoom-in fine mapping near $\bar{\Gamma}$ measured by laser-based ARPES (ii), showing nice agreement with corresponding first-principles calculations (iii,iv). The surface Dirac points (DPs) are indicated by the red arrows in (ii) and (iv). f) Band dispersion plot of 2M-WS$_2$ measured by synchrotron-based ARPES (i) and a zoom-in fine plot near $\bar{\Gamma}$ measured by laser-based ARPES (ii) along $\bar{Y} - \bar{\Gamma} - \bar{Y}$, showing consistent results with corresponding first-principles calculations (iii,iv). The surface Dirac points (DPs) are indicated by the red arrows in (ii) and (iv).

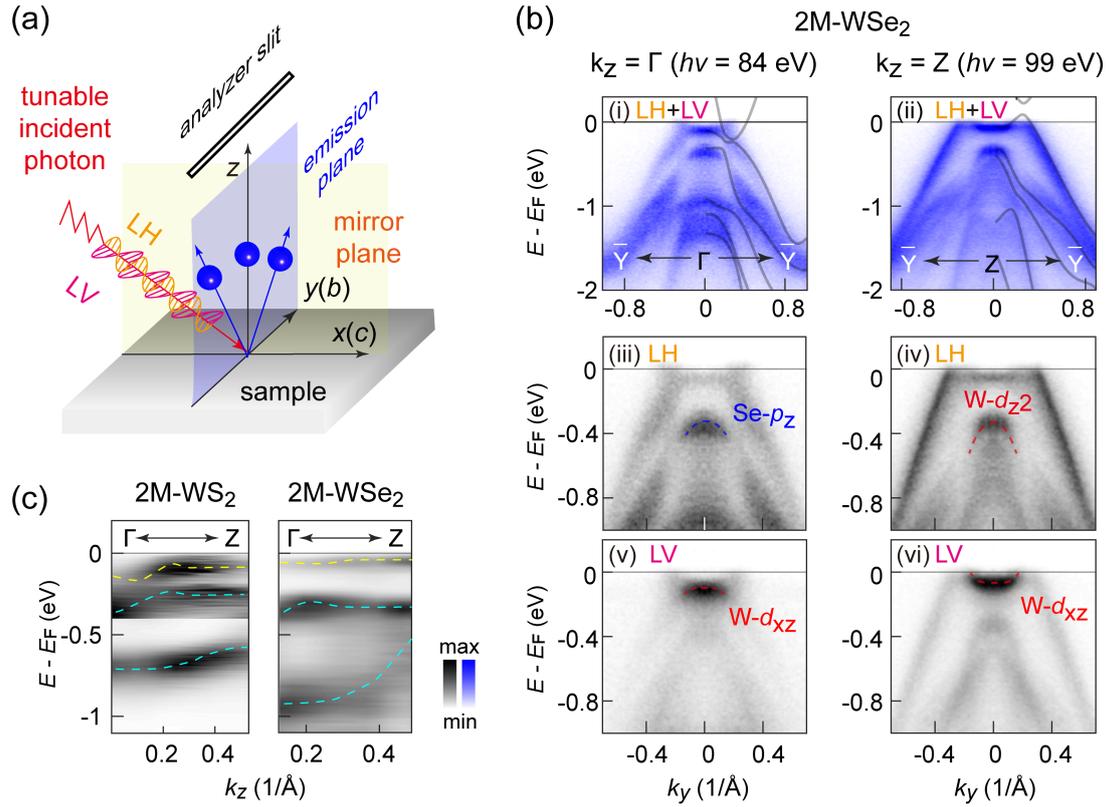

Figure 4. Photon-energy- and polarization-dependent ARPES measurements. a) Schematic polarization geometry used in the ARPES experiment, where the LH and LV-polarized light are incident on the sample parallel and perpendicular with respect to the detection plane (*xz*-plane). b)(i) and (ii) Polarization-integrated ARPES results of 2M-WSe$_2$ dispersion along (i) $\bar{Y} - \Gamma - \bar{Y}$ with a photon energy of 84 eV and (ii) $\bar{Y} - Z - \bar{Y}$ with a photon energy of 99 eV. The corresponding calculation is appended. (iii)-(vi) Polarization-dependent ARPES results of 2M-WSe$_2$ dispersion along $\bar{Y} - \Gamma - \bar{Y}$ and $\bar{Y} - Z - \bar{Y}$ near $E_F$. ARPES results are symmetrized with respect to $k_y = 0$ based on the crystalline symmetry. Blue and red dashed lines are guides to the eyes, which indicate the bands mainly contributed by specific Se and W orbitals, respectively. See more details in Supporting Information Note 3 and Note 10. c) Photon-energy-dependent ARPES measurements on the $k_z$ dispersion of 2M-WS$_2$ and 2M-WSe$_2$. The yellow and cyan dashed lines representing the three topmost valence bands are guides to the eyes.

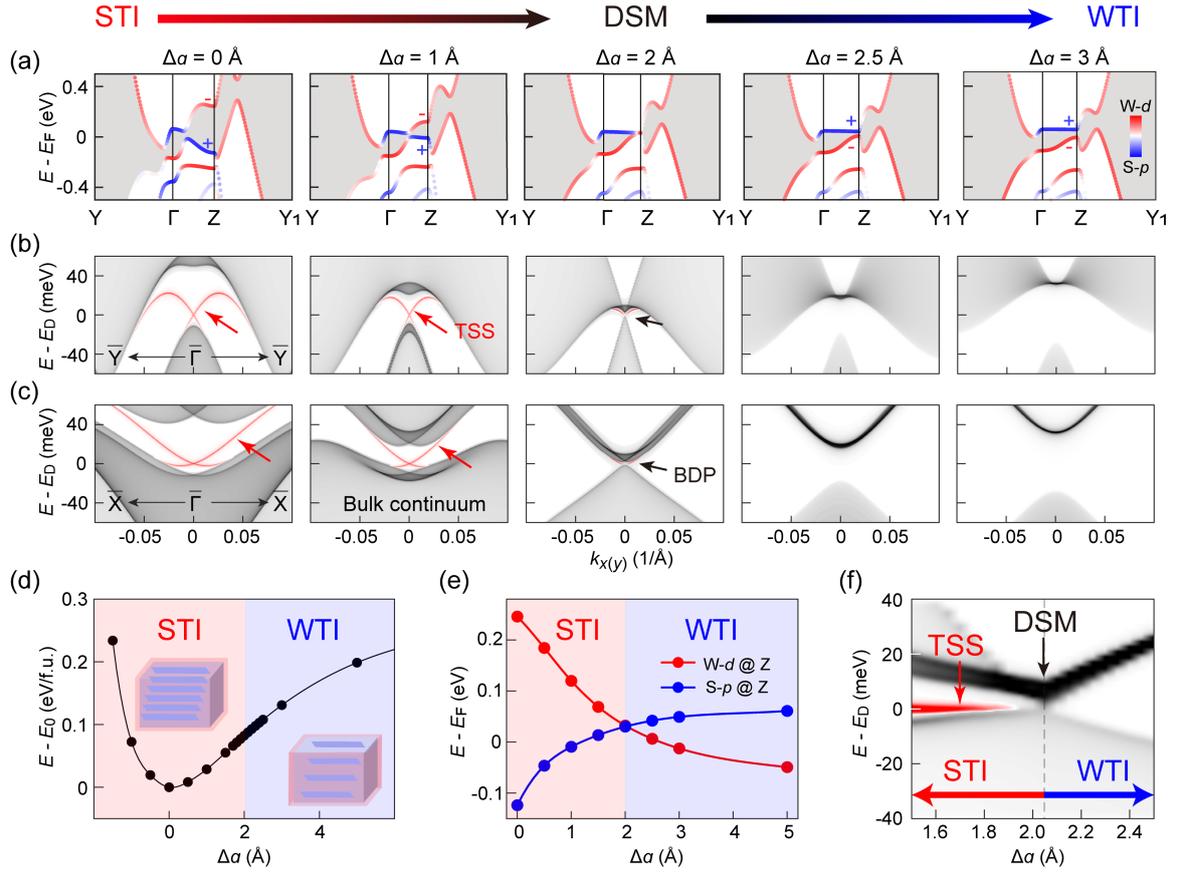

Figure 5. Calculated topological phase transition driven by the increasing interlayer distance. a) Calculated orbital-resolved electronic structure evolution of 2M-$WS_2$ under the interlayer distance indicated on the top, showing a topological phase transition from the STI phase to the WTI phase across an unstable DSM phase at the critical point. The parity eigenvalues at Z are indicated as + (even) and – (odd). STI, strong topological insulator; DSM, Dirac semimetal; WTI, weak topological insulator. b,c) Calculated surface projected band structure along $\bar{Y} - \bar{\Gamma} - \bar{Y}$ (b) and $\bar{X} - \bar{\Gamma} - \bar{X}$ (c) under the same interlayer distance as in (a). The red and black arrows indicate the TSS and the bulk Dirac point (BDP), respectively. d) Evolution of the formation energy of 2M-$WS_2$ with interlayer distance increase $\Delta a$. e) Evolution of the orbital character of the valence and conduction bands at Z, indicating orbital switching between $W$-$d$ (red) and $S$-$p$ (blue) near the critical point ($\Delta a \approx 2$Å). f) Evolution of the calculated surface projected density-of-state at $\bar{\Gamma}$ as a function of interlayer distance increase $\Delta a$, showing the TSS merges in the bulk Dirac point of the unstable DSM phase at the critical point ($\Delta a \approx 2$Å).

**References**

[1] A. H. Castro Neto, F. Guinea, N. M. R. Peres, K. S. Novoselov, A. K. Geim, *Rev. Mod. Phys.* **2009**, 81, 109.

[2] Y. Saito, T. Nojima, Y. Iwasa, *Nat. Rev. Mater.* **2017**, 2, 16094.

[3] G. Wang, A. Chernikov, M. M. Glazov, T. F. Heinz, X. Marie, T. Amand, B. Urbaszek, *Rev. Mod. Phys.* **2018**, 90, 021001.

[4] M. Gibertini, M. Koperski, A. F. Morpurgo, K. S. Novoselov, *Nat. Nanotechnol.* **2019**, 14, 408.

[5] Y. Ren, Z. Qiao, Q. Niu, *Rep. Prog. Phys.* **2016**, 79, 066501.

[6] Y. Zhang, T. R. Chang, B. Zhou, Y. T. Cui, H. Yan, Z. K. Liu, F. Schmitt, J. Lee, R. Moore, Y. L. Chen, H. Lin, H. T. Jeng, S. K. Mo, Z. Hussain, A. Bansil, Z. X. Shen, *Nat. Nanotechnol.* **2014**, 9, 111.

[7] Y. Gong, J. Guo, J. Li, K. Zhu, M. Liao, X. Liu, Q. Zhang, L. Gu, L. Tang, X. Feng, D. Zhang, W. Li, C. Song, L. Wang, P. Yu, X. Chen, Y. Wang, H. Yao, W. Duan, Y. Xu, S.-C. Zhang, X. Ma, Q.-K. Xue, K. He, *Chin. Phys. Lett.* **2019**, 36, 076801.

[8] J. H. Li, Y. Li, S. Q. Du, Z. Wang, B. L. Gu, S. C. Zhang, K. He, W. H. Duan, Y. Xu, *Sci. Adv.* **2019**, 5, eaaw5685.

[9] D. Zhang, M. Shi, T. Zhu, D. Xing, H. Zhang, J. Wang, *Phys. Rev. Lett.* **2019**, 122, 206401.

[10] M. M. Otrokov, I. I. Klimovskikh, H. Bentmann, D. Estyunin, A. Zeugner, Z. S. Aliev, S. Gass, A. U. B. Wolter, A. V. Koroleva, A. M. Shikin, M. Blanco-Rey, M. Hoffmann, I. P. Rusinov, A. Y. Vyazovskaya, S. V. Eremeev, Y. M. Koroteev, V. M. Kuznetsov, F. Freyse, J. Sanchez-Barriga, I. R. Amiraslanov, M. B. Babanly, N. T. Mamedov, N. A. Abdullayev, V. N. Zverev, A. Alfonsov, V. Kataev, B. Buechner, E. F. Schwier, S. Kumar, A. Kimura, L. Petaccia, G. Di Santo, R. C. Vidal, S. Schatz, K. Kissner, M. Uenzelmann, C. H. Min, S. Moser, T. R. F. Peixoto, F. Reinert, A. Ernst, P. M. Echenique, A. Isaeva, E. V. Chulkov, *Nature* **2019**, 576, 416.

[11] Y. J. Deng, Y. J. Yu, M. Z. Shi, Z. X. Guo, Z. H. Xu, J. Wang, X. H. Chen, Y. B. Zhang, *Science* **2020**, 367, 895.

[12] C. Liu, Y. C. Wang, H. Li, Y. Wu, Y. X. Li, J. H. Li, K. He, Y. Xu, J. S. Zhang, Y. Y. Wang, *Nat. Mater.* **2020**, 19, 522.

[13] Y. J. Chen, L. X. Xu, J. H. Li, Y. W. Li, C. F. Zhang, H. Li, Y. Wu, A. J. Liang, C. Chen, S. W. Jung, C. Cacho, H. Y. Wang, Y. H. Mao, S. Liu, M. X. Wang, Y. F. Guo, Y. Xu, Z. K. Liu, L. X. Yang, Y. L. Chen, *Phys. Rev. X* **2019**, 9, 041040.

[14] Y. Cao, V. Fatemi, A. Demir, S. Fang, S. L. Tomarken, J. Y. Luo, J. D. Sanchez-Yamagishi, K. Watanabe, T. Taniguchi, E. Kaxiras, R. C. Ashoori, P. Jarillo-Herrero, *Nature* **2018**, 556, 80.

[15] Y. Cao, V. Fatemi, S. Fang, K. Watanabe, T. Taniguchi, E. Kaxiras, P. Jarillo-Herrero, *Nature* **2018**, 556, 43.

[16] K. P. Nuckolls, M. Oh, D. Wong, B. Lian, K. Watanabe, T. Taniguchi, B. A. Bernevig, A. Yazdani, *Nature* **2020**, 588, 610.

[17] J. M. Park, Y. Cao, L.-Q. Xia, S. Sun, K. Watanabe, T. Taniguchi, P. Jarillo-Herrero, *Nat. Mater.* **2022**, 21, 877.

[18] Y. Li, S. Zhang, F. Chen, L. Wei, Z. Zhang, H. Xiao, H. Gao, M. Chen, S. Liang, D. Pei, L. Xu, K. Watanabe, T. Taniguchi, L. Yang, F. Miao, J. Liu, B. Cheng, M. Wang, Y. Chen, Z. Liu, *Adv. Mater.* **2022**, e2205996.

[19] B. A. Bernevig, S. C. Zhang, *Phys. Rev. Lett.* **2006**, 96, 106802.

[20] M. Koenig, H. Buhmann, L. W. Molenkamp, T. Hughes, C.-X. Liu, X.-L. Qi, S.-C. Zhang, *J. Phys.*

*Soc. Jpn.* **2008**, 77, 031007.

[21] A. Molle, J. Goldberger, M. Houssa, Y. Xu, S.-C. Zhang, D. Akinwande, *Nat. Mater.* **2017**, 16, 163.

[22] S. F. Wu, V. Fatemi, Q. D. Gibson, K. Watanabe, T. Taniguchi, R. J. Cava, P. Jarillo-Herrero, *Science* **2018**, 359, 76.

[23] X. F. Qian, J. W. Liu, L. Fu, J. Li, *Science* **2014**, 346, 1344.

[24] S. J. Tang, C. F. Zhang, D. Wong, Z. Pedramrazi, H. Z. Tsai, C. J. Jia, B. Moritz, M. Claassen, H. Ryu, S. Kahn, J. Jiang, H. Yan, M. Hashimoto, D. H. Lu, R. G. Moore, C. C. Hwang, C. Hwang, Z. Hussain, Y. L. Chen, M. M. Ugeda, Z. Liu, X. M. Xie, T. P. Devereaux, M. F. Crommie, S. K. Mo, Z. X. Shen, *Nat. Phys.* **2017**, 13, 683.

[25] P. Chen, W. W. Pai, Y. H. Chan, W. L. Sun, C. Z. Xu, D. S. Lin, M. Y. Chou, A. V. Fedorov, T. C. Chiang, *Nat. Commun.* **2018**, 9, 2003.

[26] Z. J. Wang, B. J. Wieder, J. Li, B. H. Yan, B. A. Bernevig, *Phys. Rev. Lett.* **2019**, 123, 186401.

[27] A. A. Soluyanov, D. Gresch, Z. J. Wang, Q. S. Wu, M. Troyer, X. Dai, B. A. Bernevig, *Nature* **2015**, 527, 495.

[28] J. Jiang, Z. K. Liu, Y. Sun, H. F. Yang, C. R. Rajamathi, Y. P. Qi, L. X. Yang, C. Chen, H. Peng, C. C. Hwang, S. Z. Sun, S. K. Mo, I. Vobornik, J. Fujii, S. S. P. Parkin, C. Felser, B. H. Yan, Y. L. Chen, *Nat. Commun.* **2017**, 8, 13973.

[29] Y. Q. Fang, J. Pan, D. Q. Zhang, D. Wang, H. T. Hirose, T. Terashima, S. Uji, Y. H. Yuan, W. Li, Z. Tian, J. M. Xue, Y. H. Ma, W. Zhao, Q. K. Xue, G. Mu, H. Zhang, F. Q. Huang, *Adv. Mater.* **2019**, 31, 1901942.

[30] Z. C. Lai, Q. Y. He, T. Ha Tran, D. V. M. Repaka, D. D. Zhou, Y. Sun, S. B. Xi, Y. X. Li, A. Chaturvedi, C. L. Tan, B. Chen, G. H. Nam, B. Li, C. Y. Ling, W. Zhai, Z. Y. Shi, D. Y. Hu, V. Sharma, Z. N. Hu, Y. Chen, Z. C. Zhang, Y. F. Yu, X. R. Wang, R. V. Ramanujan, Y. M. Ma, K. Hippalgaonkar, H. Zhang, *Nat. Mater.* **2021**, 20, 1113.

[31] L. Fu, C. L. Kane, E. J. Mele, *Phys. Rev. Lett.* **2007**, 98, 106803.

[32] S. H. Kooi, G. van Miert, C. Ortix, *Phys. Rev. B* **2018**, 98, 245102.

[33] R. Noguchi, T. Takahashi, K. Kuroda, M. Ochi, T. Shirasawa, M. Sakano, C. Bareille, M. Nakayama, M. D. Watson, K. Yaji, A. Harasawa, H. Iwasawa, P. Dudin, T. K. Kim, M. Hoesch, V. Kandyba, A. Giampietri, A. Barinov, S. Shin, R. Arita, T. Sasagawa, T. Kondo, *Nature* **2019**, 566, 518.

[34] R. Noguchi, M. Kobayashi, Z. Z. Jiang, K. Kuroda, T. Takahashi, Z. F. Xu, D. Lee, M. Hirayama, M. Ochi, T. Shirasawa, P. Zhang, C. Lin, C. Bareille, S. Sakuragi, H. Tanaka, S. Kunisada, K. Kurokawa, K. Yaji, A. Harasawa, V. Kandyba, A. Giampietri, A. Barinov, T. K. Kim, C. Cacho, M. Hashimoto, D. H. Lu, S. Shin, R. Arita, K. J. Lai, T. Sasagawa, T. Kondo, *Nat. Mater.* **2021**, 20, 473.

[35] N. Shumiya, M. S. Hossain, J. X. Yin, Z. W. Wang, M. Litskevich, C. Yoon, Y. K. Li, Y. Yang, Y. X. Jiang, G. M. Cheng, Y. C. Lin, Q. Zhang, Z. J. Cheng, T. A. Cochran, D. Multer, X. P. Yang, B. Casas, T. R. Chang, T. Neupert, Z. J. Yuan, S. Jia, H. Lin, N. Yao, L. Balicas, F. Zhang, Y. G. Yao, M. Z. Hasan, *Nat. Mater.* **2022**, 21, 1111

[36] C. Lin, M. Ochi, R. Noguchi, K. Kuroda, M. Sakoda, A. Nomura, M. Tsubota, P. Zhang, C. Bareille, K. Kurokawa, Y. Arai, K. Kawaguchi, H. Tanaka, K. Yaji, A. Harasawa, M. Hashimoto, D. H. Lu, S. Shin, R. Arita, S. Tanda, T. Kondo, *Nat. Mater.* **2021**, 20, 1093.

[37] C. Chen, A. J. Liang, S. Liu, S. M. Nie, J. W. Huang, M. X. Wang, Y. W. Li, D. Pei, H. F. Yang, H. J.

Zheng, Y. Zhang, D. H. Lu, M. Hashimoto, A. Barinov, C. Jozwiak, A. Bostwick, E. Rotenberg, X. F. Kou, L. X. Yang, Y. F. Guo, Z. J. Wang, H. T. Yuan, Z. K. Liu, Y. L. Chen, *Matter* **2020**, 3, 2055.

[38] Y. W. Li, H. J. Zheng, Y. Q. Fang, D. Q. Zhang, Y. J. Chen, C. Chen, A. J. Liang, W. J. Shi, D. Pei, L. X. Xu, S. Liu, J. Pan, D. H. Lu, M. Hashimoto, A. Barinov, S. W. Jung, C. Cacho, M. X. Wang, Y. He, L. Fu, H. J. Zhang, F. Q. Huang, L. X. Yang, Z. K. Liu, Y. L. Chen, *Nat. Commun.* **2021**, 12, 2874.

[39] C. S. Lian, C. Si, W. H. Duan, *Nano Lett.* **2021**, 21, 709.

[40] Y. H. Yuan, J. Pan, X. T. Wang, Y. Q. Fang, C. L. Song, L. L. Wang, K. He, X. C. Ma, H. J. Zhang, F. Q. Huang, W. Li, Q. K. Xue, *Nat. Phys.* **2019**, 15, 1046.

[41] Y. R. Ji, Y. B. Chu, A. M. Zhi, J. P. Tian, J. Tang, L. Liu, F. F. Wu, Y. L. Yuan, R. Yang, X. Z. Tian, D. X. Shi, X. D. Bai, W. Yang, G. Y. Zhang, *Phys. Rev. B* **2022**, 105, L161402.

[42] M. N. Ali, J. Xiong, S. Flynn, J. Tao, Q. D. Gibson, L. M. Schoop, T. Liang, N. Haldolaarachchige, M. Hirschberger, N. P. Ong, R. J. Cava, *Nature* **2014**, 514, 205.

[43] L. Fu, C. L. Kane, *Phys. Rev. B* **2007**, 76, 045302.

[44] Y. Chen, X. Gu, Y. Li, X. Du, L. Yang, Y. Chen, *Matter* **2020**.

[45] B. J. Yang, N. Nagaosa, *Nat. Commun.* **2014**, 5, 4898.

[46] S. Y. Xu, Y. Xia, L. A. Wray, S. Jia, F. Meier, J. H. Dil, J. Osterwalder, B. Slomski, A. Bansil, H. Lin, R. J. Cava, M. Z. Hasan, *Science* **2011**, 332, 560.

[47] M. Rajapakse, B. Karki, U. O. Abu, S. Pishgar, M. R. K. Musa, S. M. S. Riyadh, M. Yu, G. Sumanasekera, J. B. Jasinski, *npj 2D Mater. Appl.* **2021**, 5, 30.

[48] J. Zhou, H. W. Xu, Y. L. Shi, J. Li, *Adv. Sci.* **2021**, 8, 2003832.

[49] L. Peng, Y. Yuan, G. Li, X. Yang, J. J. Xian, C. J. Yi, Y. G. Shi, Y. S. Fu, *Nat. Commun.* **2017**, 8, 659.

[50] H. M. Weng, X. Dai, Z. Fang, *Phys. Rev. X* **2014**, 4, 011002.

[51] P. Zhang, R. Noguchi, K. Kuroda, C. Lin, K. Kawaguchi, K. Yaji, A. Harasawa, M. Lippmaa, S. M. Nie, H. M. Weng, V. Kandyba, A. Giampietri, A. Barinov, Q. Li, G. D. Gu, S. Shin, T. Kondo, *Nat. Commun.* **2021**, 12, 406.

[52] J. W. Huang, S. Li, C. Yoon, J. S. Oh, H. Wu, X. Y. Liu, N. Dhale, Y. F. Zhou, Y. C. Guo, Y. C. Zhang, M. Hashimoto, D. H. Lu, J. Denlinger, X. Q. Wang, C. N. Lau, R. J. Birgeneau, F. Zhang, B. Lv, M. Yi, *Physical Review X* **2021**, 11, 031042.

[53] M. Yang, Y. D. Liu, W. Zhou, C. Liu, D. Mu, Y. N. Liu, J. O. Wang, W. C. Hao, J. Li, J. X. Zhong, Y. Du, J. C. Zhuang, *ACS Nano* **2022**, 16, 3036.

[54] L. X. Xu, Y. Y. Y. Xia, S. Liu, Y. W. Li, L. Y. Wei, H. Y. Wang, C. W. Wang, H. F. Yang, A. J. Liang, K. Huang, T. Deng, W. Xia, X. Zhang, H. J. Zheng, Y. J. Chen, L. X. Yang, M. X. Wang, Y. F. Guo, G. Li, Z. K. Liu, Y. L. Chen, *Phys. Rev. B* **2021**, 103, L201109.

[55] J. C. Zhuang, J. Li, Y. D. Liu, D. Mu, M. Yang, Y. N. Liu, W. Zhou, W. C. Hao, J. X. Zhong, Y. Du, *ACS Nano* **2021**, 15, 14850.

[56] S. Manzeli, D. Ovchinnikov, D. Pasquier, O. V. Yazyev, A. Kis, *Nat. Rev. Mater.* **2017**, 2, 17033.

[57] M. Yankowitz, J. Jung, E. Laksono, N. Leconte, B. L. Chittari, K. Watanabe, T. Taniguchi, S. Adam, D. Graf, C. R. Dean, *Nature* **2018**, 557, 404.

[58] C. D. Zhao, X. L. Che, Y. Q. Fang, X. Y. Liu, F. Q. Huang, *J. Phys. Chem. Solids* **2021**, 149, 109789.

[59] C. Zhao, X. Che, Z. Zhang, F. Huang, *Dalton Trans.* **2021**, 50, 3862.

[60] X. L. Che, Y. J. Deng, Y. Q. Fang, J. Pan, Y. J. Yu, F. Q. Huang, *Adv. Electron. Mater* **2019**, 5, 1900462.

[61] Y. Q. Fang, Q. Dong, J. Pan, H. Y. Liu, P. Liu, Y. Y. Sun, Q. J. Li, W. Zhao, B. B. Liu, F. Q. Huang, *J. Mater. Chem. C* **2019**, 7, 8551.
[62] Y. Q. Fang, D. Wang, W. Zhao, F. Q. Huang, *EPL* **2020**, 131, 10005.
[63] G. Kresse, J. Furthmuller, *Phys. Rev. B* **1996**, 54, 11169.
[64] J. P. Perdew, K. Burke, M. Ernzerhof, *Phys. Rev. Lett.* **1996**, 77, 3865.
[65] J. C. Gao, Q. S. Wu, C. Persson, Z. J. Wang, *Computer Physics Communications* **2021**, 261, 107760.
[66] S. Grimme, J. Antony, S. Ehrlich, H. Krieg, *Journal of Chemical Physics* **2010**, 132, 154104.
[67] Q. S. Wu, S. N. Zhang, H. F. Song, M. Troyer, A. A. Soluyanov, *C Comput. Phys. Commun.* **2018**, 224, 405.
[68] A. A. Mostofi, J. R. Yates, G. Pizzi, Y. S. Lee, I. Souza, D. Vanderbilt, N. Marzari, *Comput. Phys. Commun.* **2014**, 185, 2309.
[69] N. Marzari, D. Vanderbilt, *Phys. Rev. B* **1997**, 56, 12847.
[70] I. Souza, N. Marzari, D. Vanderbilt, *Phys. Rev. B* **2002**, 65, 035109.